\documentstyle[aps,preprint]{revtex}

\newcommand{\be}{\begin{equation}}
\newcommand{\ee}{\end{equation}}
\newcommand{\bea}{\begin{eqnarray}}
\newcommand{\eea}{\end{eqnarray}}

\begin{document}
\baselineskip=12pt
\draft
\preprint{Alberta-Thy-07-94}
\title{
Dynamics of Massive Shell ejected in a Supernova Explosion}
\author{ Dar\'\i o N\'u\~nez$^{\ast}$,
H. P. de Oliveira$^{\dagger}$}
\maketitle
Theoretical Physics Institute, University of Alberta, Edmonton,
Alberta T6G
2J1, Canada. \\ e-mail nunez@phys.ualberta.ca,
oliveira@phys.ualberta.ca\\

$^{\ast}$ Instituto de Ciencias Nucleares, UNAM, \\
CU, A.P.
70-543, M\'exico, D. F. 04510,  M\'exico.\\ e-mail
nunez@roxanne.nuclecu.unam.mx\\

$^{\dagger}$Universidade do Estado do Rio de Janeiro, Instituto de
Fisica, \\
R. Sao Francisco Xavier, 524, Maracana, CEP 20550, Rio de Janeiro. \\


\begin{abstract}
{\it An expanding shell is accelerated outward by radiation from the
remnant
star and slowed by ram pressure and accretion as it plows into the
interstellar
medium. We set up the general relativistic equations of motion for
such a
shell. In the non-relativistic limit, the reduce to the standard
Ostriker-Gunn
equations. Furthermore, it is shown that the motion equation can be
integrated
in general, reducing in this way the study of  the dynamics of a
shell to a set
of four coupled first order differential equations, solvable up  to
quadrature.}
\end{abstract}
\vspace{7mm}
Subject headings: Shells, dynamics, explosion.
\vspace{7mm}
\pacs{PACS number: 97.60L}
\vfill
\eject

\centerline{1. INTRODUCTION}
\vspace{7mm}
The study of the dynamics of a supernova remnant,
a plerion,
has been greatly developed since the works of Ostriker and Gunn
(1971), where
they modeled the dynamics of this remnant as a  thin shell moving in
a
background which inside has a radiating mass and outside is dust.
Supposing
spherical symmetry, they started with the Newton's second law for
describing
the acceleration of the shell. The force terms were introduced
essentially
by hand. This is the equation  of motion:
\be
M_s{{d^2\,R}\over{dt^2}}=-G{{(M_N+{1\over 2}M_S)\,M_S}\over R^2} +
4\,\pi\,R^2\,(P_c-P_{IS}),\label{eq:o1}
\ee
where $M_S$ is the mass of the shell, $M_N$ the mass of the neutron
star, $R$
is the radius of the shell, $P_c$ is the pressure in the cavity,
related to the
radiation emitted by the star, and $P_{IS}$ is the pressure due to
the contact
with the interstellar medium, defined as
$P_{IS}=\rho_{IS}\,({{d\,R}\over{d\,t}})^2$, a friction term known as
the "snow
plow" effect (Oort 1946). The first term represents
the gravitational and self-gravitational force, the second a driving
force due
to the radiation and the third a decelerating one due to the
interaction with
the medium.

The dynamical study is completed with an equation  for the change of
the mass
of the shell (usually due only to the dust that the shell collects
while moving
in the medium) and finally an equation  for the change of the shell's
internal
energy, proportional to the radiation emitted by the star.

This system of equations has been analysed in many ways. Chevalier
(1977) and
Ostriker \& Mc Kee (1988) have analysed the kinetic equation  of
the shell to show the different stages of the evolution when each of
the terms
in turn becomes dominant, while Sato (1988a) and Sato \& Yamada
(1991) have
applied this set to the study of the SN1987A in particular. Most of
the studies
use numerical methods to obtain the change in the luminosity or in
the mass with respect to the change in the radius of the remnant.

\vspace{7mm}
\centerline {2. SHELLS IN GENERAL RELATIVITY}
\vspace{7mm}
On the other hand, the study of the dynamics of a shell separating
two
backgrounds in the context of General Relativity has developed in a
powerful
and direct formalism since the firsts works of Werner Israel (1966),
and has
been applied to cosmology, mainly to inflation  (Berezin1987), and to
modeling
the dynamics of the border between two regions in different states,
like
bubbles in water, or between two given spaces (Sato 1988b).
Nevertheless, it
has not been widely used in the study of the dynamics of the plerion.

In the context of General Relativity the equation  of motion for the
shell is
obtained from Einstein's equations and from the junction conditions
for the
metric tensor and its first derivative. We will work in spherically
symmetric spaces, described by
\be
ds^2=-e^{2\,\psi}\,f\,dv^2-2\,b\,e^{\psi}\,dr\,dv
+r^2\,d\Omega^2,\label{eq:le}
\ee
where $f$ and $\psi$ are in general functions of the null coordinate
$v$ and of
the radius $r$, and $b=\pm 1$ depending whether the null coordinate
is advanced
or retarded. The kinetic equation  for the shell  in this case are
the
following,
with our conventions following MTW (Misner, Thorne, \& Wheeler 1973),
\bea
\left[ n_\alpha \,{{\delta u^\alpha}\over{\delta \tau}} \right] &=
&{M\over
R^2} + 8\,\pi\,p,\label{eq:me1}  \\
\sigma (n_\alpha \,{{\delta u^\alpha}\over{\delta \tau}}|_+ +
n_\alpha \,{{\delta u^\alpha}\over{\delta \tau}}|_- )&= &
2\{-[T_{\alpha\,\beta}\,n^\alpha\,n^\beta] + {{p\,(F_+ + F_-)}\over
R}\},\label{eq:me2}
\eea
where $n_\alpha$ is the 4-vector normal to the shell, $u^\alpha$ is
the
4-velocity vector , $\tau$ is the proper time, $M=4\,\pi\, R^2\,
\sigma$ is the
proper mass of the shell, $\sigma$ being the surface energy density
of the
shell, and $R$ is the radius. A function in square brackets stands
for the
difference of that function in the region outside, denoted by $+$,
with the
region inside the shell, denoted by $-$, and evaluated on the shell.
Finally
$p$ is the pressure of the shell, and $F_\pm$ is given by
\be
F_\pm =\sqrt{f_\pm + {\dot R}^2},\label{eq:F}
\ee
and dot $\dot {}$ stands for differentiation with respect to the
proper
time, $\tau$. In the general relativistic context the dynamical study
is
completed by the
conservation equation, which is an equation for the rate of change of
the
proper mass
\be
\dot M + p\,\dot A =4\pi\,R^2
\,[T_{\alpha\,\beta}\,u^\alpha\,n^\beta],
\label{eq:motm}\ee
where $A=4\,\pi\,R^2$, is the area of the shell. For a derivation of
the
kinetic equations in some particular cases, see for instance (de la
Cruz \&
Israel
1967).

\vspace{7mm}
\centerline {3. INTEGRATION OF THE MOTION EQUATIONS}
\vspace{7mm}
In the present work we want to show that the formalism
of shells in General Relativity gives not only a motion equation more
general
than the Newtonian one with tiny correction whose measure is "beyond
the
present technology", but first: includes all the possible terms which
generate
the motion of the shell, no more and no less. Second: even though
there are
terms which in the Newtonian case seem to be highly dissipative, like
friction
or radiation terms, we will show that all these terms can be
described within
the formalism of General Relativity. And Third: we will show that the
motion
equation, {\it in general}, allows a first integral, which
reduces the problem to a first order dynamical set of equations. We
also
present the Newtonian limits of this integral of the motion

For the line element given by equation (\ref{eq:le}), we have that
the
4-velocity and the normal vector, already orthonormalized, at the
shell are
given by
\be
u^\alpha=(\dot v, \dot R, 0,0),\hfil  n_\alpha=b\,e^\psi(-\dot R,\dot
v,0,0)
\label{eq:vn}\ee
where $\dot v$ is related to $\dot R$ as follows
\be
\dot v={{-b\,\dot R +F}\over{f\,e^\psi}}.\label{eq:du}
\ee

The motion equation (\ref{eq:me1}) is then rewritten as
\be
[{{\ddot R + B}\over{F}}]= {M\over R^2} + 8\,\pi\,p,\label{eq:me1a}
\ee
where
\be
B={f,_r\over 2}+\psi,_r\, F +{b\over 2}\,f,_v\,e^\psi {\dot v}^2.
\ee

Now, from the definition of $F$, equation (\ref{eq:F}), we see that
\be
{{\dot R\,\ddot R}\over F}=\dot F - {{\dot f}\over {2F}},
\ee
so, after some manipulations, the motion equation  (\ref{eq:me1a})
can be
expressed as
\be
{d\over{d\tau}}([F]+{M\over R}) -{{\dot M}\over R}+[{{2\,B\,\dot R -
\dot
f}\over{2\,F}}] - 8\,\pi\,\dot R\,p= 0. \label{eq:me3}
\ee

The equation for the rate of change of the mass, equation
(\ref{eq:motm}), in
the spherical case can be expressed as
\be
{{\dot M}\over R}=[-{{f,_v\,e^\psi\,{\dot v}^2}\over 2}
+\psi,_r\,F\dot R] -
8\,\pi\,\dot R\,p,\label{eq:motm1}
\ee

Substituting this last equation  in the motion equation
(\ref{eq:me3}), we
obtain that
\bea
&{d\over{d\tau}}([F]+{M\over R}) + \nonumber &\\ & [{{2\,B\,\dot R -
\dot
f}\over{2\,F}}+ {{1}\over 2}\,f,_v\,e^\psi\,{\dot v}^2
-\psi,_r\,F\dot R]= 0.&
\eea

Finally, using the formula for $\dot v$, equation (\ref{eq:du}), can
be shown
that the expression inside the squared brackets is zero!, so
\be
{d\over{d\tau}}([F]+{M\over R})=0,
\ee
which implies that
\be
[F]+{M\over R}=0,\label{eq:fi}
\ee
the value of the integration constant, $C$, is zero, as can be shown
substituting back  $[F]+{M\over R}=C$ in the second motion equation
(\ref{eq:me2}).
In this way we have shown that equation (\ref{eq:fi}) is a first
integral of
the general motion equation  of a thin shell with spherical symmetry.

It proves convenient to introduce a function $m=m(u,r)$ such that
\be
f_\pm=1-{{2\,m_\pm}\over r},\label{eq:fc}
\ee
then, using equation (\ref{eq:F}), it is easy to show that
\be
{F_+}^2-{F_-}^2=-{{2\, m}\over R},\label{eq:F+}
\ee
where the function $m$ is defined as $m=m_+-m_-$, and in some cases
can be
identified with the gravitational mass\footnote{In the cases studied
by Israel
and de la Cruz or Chase, $m_{\pm}=m_{1,2}-\frac{e^2_{1,2}}{2\,r}$,
where
$m_{1,2}$ and $e_{1,2}$ stand for the gravitational masses  and
electrical
charges in both sides of the shell. They defined $m=m_2-m_1=constant$
as the
gravitational mass of the shell and identified this quantity as the
conserved
total energy of the shell.}. Combining the first integral equation
(\ref{eq:fi}), with equation (\ref{eq:F+}), can be obtain an
expression for $F$
on either side of the shell in terms of the proper mass and the
gravitational
mass
\be
F_\pm={m\over M}\mp {M \over {2\, R}}.\label{eq:FF}
\ee

The squared of this last equation , recalling equation (\ref{eq:fc}),
results
in
\be
{\dot R}^2=({m\over M})^2-1+{{m_++m_-}\over R}+{M^2\over
{4\,R^2}},\label{eq:lak}
\ee
which is the motion equation  obtain by Lake (1979) from the Lanczos
equation.

\vspace{7mm}
\centerline {4. CONCLUSIONS}
\vspace{7mm}
This result of being able to integrate the motion equation  in
general is quite
remarkable since it is valid for any matter distribution on either
side of the
shell, as well as for any type of shell. Furthermore, the integration
is valid
for any way in which the shell interacts with the two backgrounds!,
the only
requirements have been spherical symmetry, the fact that the
backgrounds
satisfy the Einstein's equations, and the conditions of continuity on
the
shell.

To have a better understanding of the meaning of the first integral
of the
motion equation, it might help to see its form in the Newtonian
limit. For this
purpose it proves better to take the form of the motion equation
given by
equation (\ref{eq:FF}), considering  the cases when the masses and
the
velocity, $\dot R$, are smaller than the unit and making a Taylor
expansion we
obtain
\be
m  \cong M + {1\over 2} \,M\,\dot{R^2}-\frac{M\,(m_- + \frac M2
)}{R}.
\ee
The above equation  simply states that the sum of the rest energy,
the kinetic
energy, the mutual potential energy and the gravitational self-energy
of the
shell, respectively in the $rhs$, is not constant, but varies
according with
the quantity $m$.

In this way, the dynamics of a shell moving between any two
backgrounds, with
spherical symmetry reduces to a set of first order equations, namely,
the
radial motion equation (\ref{eq:lak}), the balance equation for the
proper mass, equation (\ref{eq:motm1}), the motion equation for the
null
coordinate, equation (\ref{eq:du}), a state equation  for the shell,
giving a
relation between the pressure $p$ and the energy density, $\sigma$,
and for
specifying the kind of backgrounds between which the shell moves, we
have the
Einstein's equations, which for this spherical case are (Barrabes \&
Israel
1991)
\be
m_{,v}=4\,\pi\,r^2\,{T_v}^r; \\ m_{,r}=-4\,\pi\,r^2\,{T_v}^v; \\
\psi_{,r}=4\,\pi\,r\,T_{rr},
\ee
where $T_{\mu \,\nu}$ is the stress energy tensor, specified by the
distribution of matter and energy in the chosen background space.

Despite the above characteristic concerning the dynamics of massive
shells, it
will be worth while to write down conveniently the second order
motion equation
 in
order to obtain the Newtonian limit, and show how equation
(\ref{eq:o1}) is
recovered. After a direct calculation starting from equation
(\ref{eq:lak}), we
obtain:
\be
M\,\ddot R=-{{M\,(m_{-}+\frac{M}{2}\,F_{-})}\over {R^2}}\,+\frac{\dot
m_{-}\,M}{R\,\dot R}+\frac{\dot m\,F_{-}}{\dot R}-  \\
\frac{\dot M\,F_{-}\,F_{+}}{\dot R} \label{eq:sec}
\ee
Note that the first term in the $rhs$ is nothing else than the
relativistic
generalization of the gravitational interaction, and the remaining
terms are
related with the variation of $m_{-}$, $m$ and $M$. In fact, we have
obtained
the relativistic version of the second order equation  analysed by
Ostriker and
Gunn (1971) in studying the dynamics of remnants of supernovas
modelled by thin
spherical shells. To show in a clear way the role played by the three
last
terms of the above equation , let us consider that the shell
separates two
Vaidya spacetimes: the interior filled with outgoing radiation and
the exterior
with ingoing radiation. Thus, after taking the Newtonian limit of
equation
(\ref{eq:sec}), we arrive to the following expression:
\be
M\,\ddot R \cong -{{M\,(m_{-}+\frac{M}{2})}\over {R^2}}+8\,\pi\,R\,p+
4\,\pi\,R^2\,(q_{-}-q_{+})
\ee
where $q_{-}$ and $q_{+}$ are the energy density of the radiation at
each side
of the shell and $p$ is the pressure of the shell. Therefore, we have
recovered, in this
simple model, the terms that describe the effect of the pressure of
radiation inside the cavity and the snow plow effect (compare the
last term of
this equation  with the correspondent of the Ostriker's equation ).
Also, we
have an extra term related to the pressure of the shell, since
Ostriker has
apparently considered pressure-less shell. However, a more realistic
model is
realized in considering the exterior spacetime characterized by dust,
like in
the Friedman-Walker universe, in order to taken into account more
properly the
interstellar medium outside the shell.

We want to stress the fact that the second order equation , equation
(\ref{eq:sec}) obtained from the first order one, equation
(\ref{eq:lak}) is
completely equivalent to the two second order equations obtained from
the
Einstein's equations, equations (\ref{eq:me1}, \ref{eq:me2}) but has
a more
tractable form and the different terms are easier to identify.

Finally, if instead of the proper time of the shell, $\tau$, is used
the time
of an exterior or an interior observer, $v_+, v_-$, and then consider
the
radius of the shell as a function of $v$, $R=R(v)$, so that
$\dot R=R,_v\,\dot v$, after some manipulations can be shown that the
motion
equation  for the radius of the shell is given by
\be
R,_{v_\pm}=e^{\psi_\pm}\,{\cal A}\,[b\,{\cal A} + \eta\,({{m_g}\over
M}\mp
{M\over {2\,R}})]\,,\label{eq:ru}
\ee
with $\eta=\pm$, for the two possible solutions to the algebraic
equation  and
with
\be
{\cal A}^2=({{m_g}\over M})^2-1+{{(m_++m_-)}\over R}+{M^2\over
{4\,R^2}}.
\ee

Usually a problem could be posed as follows: given the two
backgrounds
separated by
the shell, {\it i. e.}, two spherically symmetric solutions to the
Einstein's
equations and a state equation  for the matter in the shell,
determine $R,
v_+, v_-$ and $M$ from the motion equations which are all first
order. In this
way we have a set of four first order coupled differential equations
for 4
unknown functions, so the problem is solvable up to quadrature.

Nevertheless the analytical integration might prove to be very hard
except for
the most simple cases, so numerical methods are still needed. In a
future work
(N\'u\~nez \& Oliveira 1994) we will present the analysis of this set
of
equations in various specific backgrounds.

\vspace{7mm}
\centerline {5. AKNOWLEGDMENTS}
\vspace{7mm}
It is a pleasure to thank Werner Israel for fruitful discussions and
comments
about the present work, as well as for warm hospitality. D. N. thanks
Direccion
General de Asuntos del Personal Academico, UNAM, for partial support.
H. de O.
would like to acknowledge CAPES for financial support.

\vskip7mm
\centerline{REFERENCES}

\vskip1pc
\noindent
Barrab\`es, C., \& Israel, W. 1991,  Phys. Rev D, 43, 1129

\vskip1pc
\noindent
Berezin, V. A., Kuzmin, V. A., \& Tkachev, I. I. 1987, Phys. Rev. D,
36, 2919

\vskip1pc
\noindent
Chase, J. E, 1972,  Ph. D thesis, University of Alberta, Canada.

\vskip1pc
\noindent
Chevalier, R. A. 1977, Astroph. and Space Science Lib., 66,
proceedings
(Supernovae, ed. Schramm D, Reidel Pub. Co.)

\vskip1pc
\noindent
De la Cruz, V., \& Israel, W. 1967, Il Nuovo Cimento, LI A, 3, 744

\vskip1pc
\noindent
Israel, W. 1966, Il Nuovo Cimento, 44 B, 1

\vskip1pc
\noindent
 ------. 1967, 48 B, 463

\vskip1pc
\noindent
Lake, K. 1979, Phys. Rev. D, 19, 2847

\vskip1pc
\noindent
Misner, C. W., Thorne, K. S., \& Wheeler, J. A. 1973, Gravitation,
(S. Fco. W.
H. Freemann and Co.)

\vskip1pc
\noindent
Oort, J. H. 1946, MNRA, 106, 159

\vskip1pc
\noindent
Ostriker, J. P., \& Gunn, J. E. 1971, ApJ, 164, L95

\vskip1pc
\noindent
Ostriker, J. P., \& McKee, C. F. 1988, Rev. Mod. Phys., 60, 1

\vskip1pc
\noindent
Sato, H. 1988a, Prog. of Theo. Phys., 80, 96

\vskip1pc
\noindent
Sato, H. 1988b, Prog. of Theo. Phys., 80, 96

\vskip1pc
\noindent
Sato, H., \& Yamada, Y. 1991, Prog. of Theo. Phys., 85, 541

\vskip1pc
\noindent
N\'u\~nez, D., \& de Oliveira, H. P., in preparation

\end{document}